\documentclass{llncs}

\usepackage{amsmath}
\usepackage{amsfonts}
\usepackage{amssymb}
\usepackage{color}
\usepackage{graphicx}
\usepackage{bm, tabularx}
\usepackage[boxed, ruled, vlined]{algorithm2e}

\newcommand{\E}{\mbox{E}}

\newcommand{\Cov}{\mbox{Cov}}

\makeatletter
\renewcommand{\@Opargbegintheorem}[4]{%
  #4\trivlist\item[\hskip\labelsep{#3#2\@thmcounterend}]}
\makeatother

\title{On the number of signals in multivariate time series}
\titlerunning{Estimation of the noise dimensionality}

\author{Markus Matilainen\inst{1} \and Klaus Nordhausen\inst{2} \and Joni Virta\inst{1}\inst{3}}

\institute{University of Turku, Finland
\and Vienna University of Technology, Austria
\and Aalto University, Finland}

\begin{document}

\maketitle

\begin{abstract}
We assume a second-order source separation model where the observed multivariate time series is a linear mixture of latent, temporally uncorrelated time series with some components pure white noise. To avoid the modelling of noise, we extract the non-noise latent components using some standard method, allowing the modelling of the extracted univariate time series individually. An important question is the determination of which of the latent components are of interest in modelling and which can be considered as noise. Bootstrap-based methods have recently been used in determining the latent dimension in various methods of unsupervised and supervised dimension reduction and we propose a set of similar estimation strategies for second-order stationary time series. Simulation studies and a sound wave example are used to show the method's effectiveness.
\end{abstract}

\section{Time series modelling via blind source separation}\label{sec:intro}

Consider a multivariate time series $\bm x_t = (x_{1t}, \ldots , x_{pt})^\top \in \mathbb{R}^p$, $t \in \{ 1, \ldots , T \}$, commonly encountered in contemporary applications in the form of e.g. climate, financial, EEG, MEG of fMRI-data \cite{TangSutherlandMcKinney2005}. Naturally, in each of these cases the series can have dependency both within and between the individual series and it is this richness of structure that sets multivariate time series analysis apart from its univariate counterpart. Needless to say, the added complexity comes with a price: already in the simplest first-order vector autoregressive VAR$(1)$-model \cite{Lutkepohl2005}, where each time point linearly depends on the values of the previous time only, it takes a total of $2 p^2$ parameters to describe the full covariance structure of the model, and with any more sophisticated models the number of parameters inflates even further. The problem with modelling is further amplified when the dimensionality $p$ is large: as multivariate data often contains varying quantities of redundancy and noise some of the model parameters are actually used to model them while in reality we could resort to a simpler model.

A simultaneous solution to both previous problems is given by (linear) \textit{blind source separation} (BSS) \cite{ComonJutten2010}
. In our time series context, we assume in BSS that the observed series $\bm x_t$ is an invertible mixture of some latent series $\bm z_t$ with a simpler dependency structure, i.e.
\begin{align}\label{eq:bss}
\bm x_t = \bm \mu + \bm \Omega \bm z_t, \quad t \in \{ 1, \ldots , T \},
\end{align}
where $\bm \mu \in \mathbb{R}^p$ is the location vector and the \textit{mixing matrix} $\bm \Omega \in \mathbb{R}^{p \times p}$ is invertible. Furthermore, $\bm z_t$ is usually assumed to be weak second-order stationary and its component series temporally uncorrelated,
\begin{align*}
\E ((\bm z_t - E(\bm z_t)) (\bm z_{t + \tau} - E(\bm z_t))^{\top})
= \bm \Lambda_{\tau}
\mbox{ is diagonal for all lags } \tau \in \mathbb{Z}_+.
\end{align*}
The assumption on stationarity further allows us to fix $\E (\bm z_t) = \bm 0, \Cov (\bm z_t) = \bm I_p$ as the two moments are in \eqref{eq:bss} confounded with $\bm \mu $ and $\bm \Omega$, respectively. The BSS model \eqref{eq:bss} equipped with the previous assumptions is commonly known as the \textit{second order separation} (SOS) model \cite{ComonJutten2010}.

Measurement error and noise are commonly included in the model \eqref{eq:bss} additively, as $\bm x_t = \bm \mu + \bm \Omega \bm z_t +  \bm \epsilon_t$ where $\bm \epsilon_t \in \mathbb{R}^p$ is a white noise vector \cite{ComonJutten2010} representing the two sources of external variation. However, as in this case all estimates of the signals will always be distorted by some noise, we work in the following  with the contrasting idea that the noise is not an external but an internal part of the model. That is, we assume that the latent series can be partitioned as $\bm z_t = (\bm s_t^\top, \bm w_t^\top)^\top$ where $\bm w_t \in \mathbb{R}^{p - k}$ is white noise and the sources of interest (``signals'') in $\bm s_t \in \mathbb{R}^k$ contain all the time dependency manifested in $\bm x_t$. Similar models (with different definitions of ``noise'') have been previously used in the context of both unsupervised and supervised dimension reduction in e.g. \cite{BlanchardKawanabeSugiyamaSpokoinyMuller2006,NordhausenOjaTyler2016,NordhausenOjaTylerVirta2017,MatilainenCrouxNordhausenOja2017}. Compared to the additive noise model the proposed one makes the modelling and predicting of $\bm x_t$ particularly simple, the process consisting of four steps: estimate the latent series $\bm z_t$ using some standard method, identify the $p - k$ white noise series among $\bm z_t$ and discard them, model the remaining $k$ temporally uncorrelated signal series individually, and finally, back-transform the model to the original scale. This recipe avoids both of the previous problems affecting multivariate time series models: the number of parameters is kept in control as instead of modelling a full $p$-variate time series we model $k$ univariate time series, and the modelling of noise is averted as we discard it prior to the modelling step.

However, the second of the four steps, the estimation of the dimensionality $k$, is often heavily overlooked in similar contexts in the literature. BSS as a solution to the modelling problem can be seen to have succeeded only partially if our estimate $d$ of $k$ is inconsistent: on one hand, having $d > k$ means the we model noise in the third step, further biasing any predictions made with the model later, and on the other hand, having $d < k$ means that not all of the signal gets captured by the model and we have voluntarily discarded information. The problem is similar to that of selecting the number of principal components in principal component analysis where na\"ive descriptive tools such as the scree plot or the Kaiser rule \cite{Jolliffe2002} are commonly used. 
To solve the problem in our context, we propose a semi-parametric, bootstrap-based strategy for estimating $k$.

\section{Two SOS methods and test statistics}

To motivate our approach we next go through the steps taken in the two most popular SOS methods, AMUSE (algorithm for multiple signals extraction) \cite{TongSoonHuangLiu1990} and SOBI (second order blind identification) \cite{BelouchraniAbedMeraimCardosoMoulines1997}. Also, without loss of generality, we assume that all our series are centered, i.e. $\bm \mu = \bm 0$. AMUSE and SOBI both assume the model \eqref{eq:bss} and the assumptions following it. We denote the lag-$\tau$ autocovariance matrix of the series $\textbf{x}_t$ by $\bm \Sigma _\tau (\textbf{x}_t) = \E ( \bm x _t \bm x _{t + \tau}^\top ) $, the choice $\tau = 0$ giving the marginal covariance matrix of the series.

The usual starting point in BSS is whitening the data: we estimate the marginal covariance matrix $\bm \Sigma _0 (\textbf{x}_t)$ and standardize the series using its (unique symmetric) inverse root $\bm \Sigma _0 (\bm x _t)^{-1/2}$. This yields us the standardized series $\bm x _t^{st} = \bm \Sigma _0 (\bm x _t)^{-1/2} \bm x _t$ with the property that $\bm \Sigma_0 ( \bm x _t^{st} ) = \bm I _p$. Some algebra reveals the importance of the standardization for the BSS model: the standardized series satisfies $\bm x _t^{st} = \bm U \bm z_t$ for some unknown orthogonal matrix $\bm U \in \mathbb{R}^{p \times p}$ \cite{CardosoSouloumiac1993,MiettinenTaskinenNordhausenOja2015}.

This insight instantly suggests using the eigendecompositions of the autocovariance matrices to recover the missing matrix $\bm U$. Following our assumptions, for any fixed lag $\tau_0 > 0$ we have $\bm \Sigma _{\tau_0} ( \bm x _t^{st} ) = \bm U \bm \Lambda _{\tau_0} \bm U ^\top$ where $\bm \Lambda _{\tau_0}$ is diagonal. The diagonal elements of $\bm \Lambda _{\tau_0}$ contain the marginal $\tau_0$th autocovariances of the latent series and for the white noise series $\bm w _t$ they naturally equal $0$. Thus, assuming that all the $k$ signal series correspond to distinct, non-zero eigenvalues, the related eigenvectors $\bm U _1 \in \mathbb{R}^{p \times k}$ can be identified up to sign and order, and finally, we obtain the signal series $\bm s _t$ via the transformation $\bm x _t \mapsto \bm U _1^\top \bm \Sigma _0 (\bm x _t)^{-1/2} \bm x _t$, yielding the AMUSE-solution with lag $\tau_0$. In practice the time series of interest are selected from the estimated $p$ latent series by inspecting the diagonal values of the estimated $\hat{ \bm \Lambda }_{\tau_0}^2$, where the squaring is used simply for convenience to order the components in a decreasing order of interestingness. The noise components can now be identified as being the last $p - d$ components that have ``small enough'' eigenvalues, $d = 0, \ldots , p$, the key question then being what actually is small enough. An equivalent formulation for the problem can be stated via the running means $m_{p - d}$ of the last $p - d$ squared eigenvalues by asking for which $d$ the estimate $\hat{m}_{p - d}$ is ``too large''. This prompts to use $\hat{m}_{p - d}$ as a test statistic for testing the null hypothesis,
\[
H_{0,d}: \mbox{The last $p - d$ latent series are white noise}.
\]
For a fixed $d$, if the observed value of $\hat{m}_{p - d}$ exceeds some pre-defined critical value we conclude that the result is too unlikely to have been originated under the null hypothesis and infer that the number of signal components is larger than $d$. Chaining together tests for several null hypotheses $H_{0,d_1}, H_{0,d_2}, \ldots$ then allows us to pinpoint the true value $d = k$. However, obtaining the distribution of our test statistic under the null hypothesis is a highly non-trivial task under the general SOS-model, and we thus resort to the bootstrap \cite{efron1979bootstrap} to obtain the quantiles, the next section detailing several bootstrapping strategies we can use to accurately replicate the null distribution.

AMUSE already gives us a reasonable starting point for devising a test statistic for the signal dimensionality, but suffers from a clear drawback: the signal components must all have non-zero $\tau_0$th autocovariances in order to be distinguished from the noise (to be distinguishable from each other the signal autocovariances also need to be mutually distinct but that is irrelevant with respect to our current problem of separating the \textit{noise subspace} from the \textit{signal subspace} as a whole). In practice this necessitates a careful choosing of the single lag $\tau_0$, possibly using some expert knowledge on the phenomenon at hand. Such inconvenience is avoided with our preferred SOS-method, SOBI. In SOBI one instead chooses a set of lags, $\mathcal{T}_0$, and \textit{jointly diagonalizes} all $| \mathcal{T}_0 |$ autocovariance matrices of the standardized series corresponding to the lags (with $| \mathcal{T}_0 | = 1$ we revert back to AMUSE). The joint diagonalization is captured by the optimization problem
\begin{align*}
\bm U ^\top = \underset{ \bm V ^\top \bm V = \bm I _p}{\mbox{argmax}} \sum_{\tau \in \mathcal{T}_0} \left\| \mbox{diag} \left(\bm V ^\top \bm \Sigma_{\tau}\left(\bm x^{st}_t \right) \bm V \right) \right\|_F^2,
\end{align*}
commonly solved using the Jacobi rotation algorithm \cite{Clarkson1988}. For two latent components to be mutually distinguishable by the joint diagonalization it is sufficient that the corresponding marginal autocovariances differ for some lag in $\mathcal{T}_0$ \cite{BelouchraniAbedMeraimCardosoMoulines1997}. In particular, we can distinguish the noise subspace from the signal subspace if all signal series exhibit autocorrelation for at least one lag in $\mathcal{T}_0$ (which can be a different lag for different signals), prompting us to choose a relatively large set of lags, $\mathcal{T}_0 = \{1, \ldots , 12\}$ being a common choice. Thus, a natural test statistic for the null hypothesis $H_{0,d}$ is again obtained by considering ``eigenvalues'', the diagonal elements of the estimated as-diagonal-as-possible matrices $\bm \Lambda _\tau = \mathrm{diag}( \bm U ^\top \bm \Sigma_{\tau}\left(\bm x^{st}_t \right) \bm U )$. Ordering the sums of the squared elements in decreasing order, the running means $\hat{m}_{p - d}$ of the last $p - d$ components of the sample estimate of $\sum_{\tau \in \mathcal{T}_0} \bm \Lambda _\tau^2$ will be ``small'' for large enough values of $d$ and their null distributions can be used to find the value of $d$ where $\hat{m}_{p - d}$ is too large to have originated under the null hypothesis, again allowing us to identify the correct dimensionality.

\section{Bootstrap tests for the white noise dimension}

Bootstrap-based methods have recently been used in determining the noise subspace dimension for principal component analysis (PCA), independent component analysis (ICA) and sliced inverse regression (SIR) in \cite{NordhausenOjaTyler2016} and for non-Gaussian component analysis in \cite{NordhausenOjaTylerVirta2017}. As an alternative testing method both works also discuss tests that are based on limiting behaviors of certain functions of the noise eigenvalues. Such asymptotic procedures are indeed efficient when the sample size is high and could certainly be considered in our context as well, if not for the general difficulty of obtaining limiting results for time series models (see however the limiting behaviour of AMUSE and SOBI for linear processes in \cite{MiettinenNordhausenOjaTaskinen2012,MiettinenNordhausenOjaTaskinen2014b,IllnerMiettinenFuchsTaskinenNordhausenOjaTheis2015,MiettinenIllnerNordhausenOjaTaskinenTheis2016,TaskinenMiettinenNordhausen2016}). As such we leave the development of asymptotic testing procedures to a subsequent work and proceed now with bootstrapping tests
.

Assume a time series coming from the model \eqref{eq:bss}, fix a candidate for the signal dimension $d$ and let $\hat{m}_{p - d}$ be the test statistic of the previous section, the mean of the last $p - d$ squared eigenvalues produced by either AMUSE or SOBI (mean of the sums of the squared ``eigenvalues'' in the case of SOBI). To test the null hypothesis $H_{0,d}$ we need a way to generate samples from the distribution of the model \eqref{eq:bss} under the null hypothesis. We will consider four different strategies where we always leave the signal part untouched and take bootstrap samples of the noise part under the current null hypothesis, denoted $z^*_{i,s}$ where $i = d+1, \dots, p$ denotes the component and $s=1,\ldots, T$ the time point.

\paragraph{Parametric bootstrap:}
The most widely used assumption about the white noise is that it is Gaussian, making all noise features independently and identically $N(0,1)$-distributed. The bootstrap samples are then
\[
z^*_{i,s} \sim N(0,1).
\]
Naturally, the parametric bootstrap makes the strongest assumptions, in this case that (i) the noise processes are independent, (ii) within a noise process the time points are serially independent and (iii) the noise is Gaussian. Using next non-parametric bootstrap these assumptions can be relaxed in different ways.


\paragraph{Non-parametric bootstrap I:}
First we relax the distributional assumption while keeping assumptions (i) and (ii), and assume only that the noise distribution is for all noise components the same but not necessarily Gaussian. Then all $(p - k) \times T$ elements in the noise part are iid samples from the same distribution and we can use use the combined sample to estimate the empirical distribution function (ecdf) and to sample $(p - d) \times T$ elements from it. Thus
\[
z^*_{i,s} \sim \mbox{ecdf}\{ ( \hat{\bm{z}}_{d+1}^\top, \ldots , \hat{\bm{z}}_{p}^\top)^\top \}, \ i=d+1,\ldots,p, \ s=1,\ldots ,T,
\]
where $\hat{\bm{z}}_{j}$ is the $T$-vector of the estimated $j$th latent series and $\mbox{ecdf}\{\bm x\}$ denotes the ecdf of the samples in $\bm x$.


\paragraph{Non-parametric bootstrap II:}

Another way to relax the third assumption is to keep assumptions (i) and (ii) but assume that each process has a possibly different standardized distribution. In that case each noise series should be bootstrapped individually and independently from the others. Therefore using this strategy the bootstrap samples are obtained as
\[
z^*_{i,s} \sim \mbox{ecdf}\{ \hat{\bm{z}}^\top_{i} \}, \ i=d+1,\ldots,p, \ s=1,\ldots ,T.
\]

\paragraph{Non-parametric bootstrap III:}

The last approach considered relaxes also the independence between the noise processes and just requires that they are uncorrelated and serially independent. Hence the ecdf is now multivariate and a bootstrap sample of vectors is obtained as
\[
\bm{z}^*_{n, s} \sim \mbox{ecdf} \{ \hat{\bm{z}}_{n, 1}, \ldots, \hat{\bm{z}}_{n, T} \}, \ s=1,\ldots ,T,
\]
where $\bm z ^*_{n, s} = (z^*_{d+1,s}, \ldots, z^*_{p,s})^\top$ and $\hat{\bm z} _{n, t} = (\hat{z}_{d+1,t}, \ldots, \hat{z}_{p,t})^\top$.

In Algorithm \ref{alg::boot} we describe the entire testing procedure for $H_{0,d}$ using SOBI  (where the version for AMUSE is obtained by using only a single lag).

\begin{algorithm}[H]\label{alg::boot}
Set the proposed dimension $d$\;
Set the number of resamples $R$\;
Set the observed sample $\bm{X}_i$\;
Estimate the SOBI-solution for $\bm{X}_i$: $\hat{\bm{U}}^\top \hat{\bm{\Sigma}}_0^{-1/2}$, $\hat{m}_{p - d}$\;
\For{$i \in \{ 1, \ldots , R \}$}{
  $\bm{Z}^*_i \leftarrow$ bootstrap the last $p - d$ series of $\hat{\bm{Z}}_i = \hat{\bm{U}}^\top \hat{\bm{\Sigma}}_0^{-1/2} \bm{X}_i$\;
  $\bm{X}^*_i \leftarrow \hat{\bm{\Sigma}}_0^{1/2} \hat{\bm{U}} \bm{Z}^*_i $\;
  Estimate the SOBI-solution for $\bm{X}^*_i$: $\hat{m}^*_{p - d}$\;
}
 Return the $p$-value: $[\#(\hat{m}^*_{p - d} \geq \hat{m}_{p - d}) + 1]/(R + 1)$\;
 \caption{Testing $H_{0,d}$}
\end{algorithm}

\vspace{0.2cm}

The addition of one in both the numerator and denominator of the $p$-value is a commonly used ``correction'' to avoid the event of obtaining a zero $p$-value. For some other guidelines concerning bootstrap hypothesis testing, see \cite{hall1991two}.

The procedure above tests only for a specific value of the signal/noise dimension. To obtain an estimate for the dimension, the changing point from rejection to acceptance of the sequence of null hypotheses is of interest. For that the tests have to be applied sequentially and different strategies are possible. For example, one can start with the assumption that all components are noise and then increase successively the hypothetical signal dimension until for the first time the null hypothesis cannot be rejected or one can start with the hypothesis of a single noise component and increase the noise dimension until the first time the null hypothesis is rejected. Another possibility is to use some divide-and-conquer strategy. Comparing different estimation strategies is however beyond the scope of this paper and will be explored in a future work. The following simulation study focuses on validating the bootstrap hypothesis tests as suggested above.


\section{Simulations}\label{sec:simulations}

In order to assess the performance of the bootstrap tests, we conducted a simulation study with three different settings using 5-dimensional time series. 
The first two are taken as ARMA-processes: $z_1 \sim ARMA(2,1)$ with parameters $\phi_1 = 0.5$, $\phi_2 = 0.2$ and $\theta_1 = 0.5$ and $z_2 \sim MA(5)$ with the parameter vector $\bm{\theta} = (-0.4, 0.6, -0.3, 0.1, -0.3)$. The final three series are noise with the following distributions in the different settings: Setting 1: $z_3, z_4, z_5 \sim  N(0,1)$; Setting 2: $(z_3, z_4, z_5) \sim  \bm t _5$; Setting 3: $z_3 \sim N(0,1), z_4 \sim t_5$ and $z_5 \sim U(-\sqrt{3},\sqrt{3})$.
%
%

In all settings the signal subspace has the true dimension $k = 2$. Setting 1 is possibly the most natural one, in Setting 2 the noise has a spherical 3-variate $\bm t _5$-distribution which means that there is some dependence among the components and in Setting 3 the noise components are independent but have different marginal distributions. As a mixing matrix we used a random matrix $\bm \Omega$, where the elements of the matrix were drawn randomly from the standard normal distribution.
Next the bootstrap $p$-values based on $M = 200$ and $500$ bootstrap samples were calculated and the procedure was repeated 2000 times.
We used the time series lengths $T = 200, 500, 2000$ and $5000$ and AMUSE with lag $1$ and SOBI with lags $1, \ldots, 12$. Tables \ref{table::UN} -- \ref{table::UM} show the proportions of rejections at the $\alpha$-level 0.05 based on 2000 repetitions for hypotheses $H_{0,1}$, $H_{0,2}$ (the true value which should be the first test we do not reject) and $H_{0,3}$ for each combination of settings and methods with $M = 200$. The results based on $M = 500$ gave very similar results and were thus omitted from the tables.


\begin{table}[t]
\centering
\caption{The rejection rates of the hypotheses in Setting 1 over 200 bootstrap samples and 2000 repetitions.}\label{table::UN}
\begin{tabular}{p{1cm}p{3cm}p{0.1cm}p{1cm}p{1cm}p{1cm}p{0.1cm}p{1cm}p{1cm}p{1cm}}
\hline
 & & & & AMUSE & & & & SOBI & \\
 \cline{4-6}\cline{8-10}
n & Booststrap method & & $H_{0,1}$ & $H_{0,2}$ & $H_{0,3}$ & & $H_{0,1}$ & $H_{0,2}$ & $H_{0,3}$\\
\cline{1-2}\cline{4-6}\cline{8-10}
200  & parametric & & 0.998 & 0.042 & 0.004 & & 1.000 & 0.049 & 0.006 \\
200  & non-parametric I & & 0.998 & 0.042 & 0.006 & & 0.998 & 0.047 & 0.008 \\
200  & non-parametric II & & 0.998 & 0.048 & 0.006 & & 1.000 & 0.046 & 0.005 \\
200  & non-parametric III & & 0.999 & 0.046 & 0.005 & & 1.000 & 0.052 & 0.005 \\
\cline{1-2}\cline{4-6}\cline{8-10}
500  & parametric & & 1.000 & 0.047 & 0.008 & & 1.000 & 0.052 & 0.010 \\
500  & non-parametric I & & 1.000 & 0.043 & 0.007 & & 1.000 & 0.047 & 0.008 \\
500  & non-parametric II & & 1.000 & 0.046 & 0.010 & & 1.000 & 0.050 & 0.010 \\
500  & non-parametric III & & 1.000 & 0.045 & 0.010 & & 1.000 & 0.054 & 0.008 \\
\cline{1-2}\cline{4-6}\cline{8-10}
2000  & parametric & & 1.000 & 0.053 & 0.008 & & 1.000 & 0.048 & 0.007 \\
2000  & non-parametric I & & 1.000 & 0.042 & 0.006 & & 1.000 & 0.057 & 0.007 \\
2000  & non-parametric II & & 1.000 & 0.052 & 0.006 & & 1.000 & 0.050 & 0.008 \\
2000  & non-parametric III & & 1.000 & 0.052 & 0.008 & & 1.000 & 0.048 & 0.008 \\
\cline{1-2}\cline{4-6}\cline{8-10}
5000  & parametric & & 1.000 & 0.052 & 0.008 & & 1.000 & 0.053 & 0.006 \\
5000  & non-parametric I & & 1.000 & 0.050 & 0.009 & & 1.000 & 0.054 & 0.006 \\
5000  & non-parametric II & & 1.000 & 0.054 & 0.010 & & 1.000 & 0.050 & 0.007 \\
5000  & non-parametric III & & 1.000 & 0.052 & 0.007 & & 1.000 & 0.050 & 0.006 \\
\hline
\end{tabular}
\end{table}


\begin{table}[t]
\centering
\caption{The rejection rates of the hypotheses in Setting 2 over 200 bootstrap samples and 2000 repetitions.}\label{table::UT}
\begin{tabular}{p{1cm}p{3cm}p{0.1cm}p{1cm}p{1cm}p{1cm}p{0.1cm}p{1cm}p{1cm}p{1cm}}
\hline
 & & & & AMUSE & & & & SOBI & \\
 \cline{4-6}\cline{8-10}
n & Booststrap method & & $H_{0,1}$ & $H_{0,2}$ & $H_{0,3}$ & & $H_{0,1}$ & $H_{0,2}$ & $H_{0,3}$\\
\cline{1-2}\cline{4-6}\cline{8-10}
200  & parametric & & 1.000 & 0.046 & 0.007 & & 1.000 & 0.052 & 0.008 \\
200  & non-parametric I & & 0.998 & 0.045 & 0.006 & & 1.000 & 0.030 & 0.004 \\
200  & non-parametric II & & 1.000 & 0.046 & 0.006 & & 1.000 & 0.048 & 0.010 \\
200  & non-parametric III & & 1.000 & 0.044 & 0.006 & & 0.999 & 0.051 & 0.012 \\
\cline{1-2}\cline{4-6}\cline{8-10}
500  & parametric & & 1.000 & 0.052 & 0.008 & & 1.000 & 0.043 & 0.007 \\
500  & non-parametric I & & 1.000 & 0.050 & 0.005 & & 1.000 & 0.046 & 0.006 \\
500  & non-parametric II & & 1.000 & 0.046 & 0.006 & & 1.000 & 0.044 & 0.006 \\
500  & non-parametric III & & 1.000 & 0.051 & 0.006 & & 1.000 & 0.046 & 0.008 \\
\cline{1-2}\cline{4-6}\cline{8-10}
2000  & parametric & & 1.000 & 0.042 & 0.003 & & 1.000 & 0.047 & 0.007 \\
2000  & non-parametric I & & 1.000 & 0.044 & 0.005 & & 1.000 & 0.051 & 0.006 \\
2000  & non-parametric II & & 1.000 & 0.048 & 0.002 & & 1.000 & 0.047 & 0.010 \\
2000  & non-parametric III & & 1.000 & 0.044 & 0.003 & & 1.000 & 0.045 & 0.008 \\
\cline{1-2}\cline{4-6}\cline{8-10}
5000  & parametric & & 1.000 & 0.050 & 0.008 & & 1.000 & 0.047 & 0.009 \\
5000  & non-parametric I & & 1.000 & 0.068 & 0.010 & & 1.000 & 0.055 & 0.006 \\
5000  & non-parametric II & & 1.000 & 0.049 & 0.005 & & 1.000 & 0.048 & 0.008 \\
5000  & non-parametric III & & 1.000 & 0.049 & 0.007 & & 1.000 & 0.046 & 0.006 \\
\hline
\end{tabular}
\end{table}


\begin{table}[t]
\centering
\caption{The rejection rates of the hypotheses in Setting 3 over 200 bootstrap samples and 2000 repetitions.}\label{table::UM}
\begin{tabular}{p{1cm}p{3cm}p{0.1cm}p{1cm}p{1cm}p{1cm}p{0.1cm}p{1cm}p{1cm}p{1cm}}
\hline
 & & & & AMUSE & & & & SOBI & \\
 \cline{4-6}\cline{8-10}
n & Booststrap method & & $H_{0,1}$ & $H_{0,2}$ & $H_{0,3}$ & & $H_{0,1}$ & $H_{0,2}$ & $H_{0,3}$\\
\cline{1-2}\cline{4-6}\cline{8-10}
200  & parametric & & 0.999 & 0.044 & 0.006 & & 0.998 & 0.051 & 0.008 \\
200  & non-parametric I & & 0.999 & 0.063 & 0.008 & & 0.999 & 0.047 & 0.009 \\
200  & non-parametric II & & 0.998 & 0.044 & 0.006 & & 0.998 & 0.047 & 0.008 \\
200  & non-parametric III & & 0.998 & 0.044 & 0.007 & & 0.999 & 0.046 & 0.007 \\
\cline{1-2}\cline{4-6}\cline{8-10}
500  & parametric & & 1.000 & 0.045 & 0.008 & & 1.000 & 0.054 & 0.006 \\
500  & non-parametric I & & 1.000 & 0.052 & 0.005 & & 1.000 & 0.047 & 0.006 \\
500  & non-parametric II & & 1.000 & 0.044 & 0.005 & & 1.000 & 0.052 & 0.009 \\
500  & non-parametric III & & 1.000 & 0.051 & 0.006 & & 1.000 & 0.056 & 0.006 \\
\cline{1-2}\cline{4-6}\cline{8-10}
2000  & parametric & & 1.000 & 0.050 & 0.004 & & 1.000 & 0.051 & 0.007 \\
2000  & non-parametric I & & 1.000 & 0.050 & 0.006 & & 1.000 & 0.060 & 0.006 \\
2000  & non-parametric II & & 1.000 & 0.050 & 0.006 & & 1.000 & 0.052 & 0.009 \\
2000  & non-parametric III & & 1.000 & 0.049 & 0.003 & & 1.000 & 0.046 & 0.008 \\
\cline{1-2}\cline{4-6}\cline{8-10}
5000  & parametric & & 1.000 & 0.060 & 0.007 & & 1.000 & 0.046 & 0.009 \\
5000  & non-parametric I & & 1.000 & 0.053 & 0.008 & & 1.000 & 0.053 & 0.009 \\
5000  & non-parametric II & & 1.000 & 0.058 & 0.004 & & 1.000 & 0.045 & 0.009 \\
5000  & non-parametric III & & 1.000 & 0.058 & 0.006 & & 1.000 & 0.046 & 0.006 \\
\hline
\end{tabular}
\end{table}

Based on the simulation results one can conclude that all the tests had quite good power  and successfully detected if there were non-noise components among the hypothetical noise part. An interesting feature is that the parametric bootstrap test seems quite robust -- it works also in Settings 2 and 3 where the data were generated using other noise processes. The non-parametric bootstrap test I, which is the closest to the parametric one, seems however to be the least effective of the non-parametric bootstrap tests. As the non-parametric bootstrap test III is valid in all three settings, and the other tests do not gain much in the settings they were designed for, this test might be the best choice in practise.
As the differences between using AMUSE or SOBI seem minor here, we advocate SOBI for practical applications as it is usually preferable over AMUSE and most likely estimates the signals better.

\section{Sound example}

\renewcommand{\thefootnote}{\arabic{footnote}}

To evaluate the method in practice we used it to estimate the dimension of a set of sound recordings mixed with noise. The signal part $\bm{s}_t$ was $3$-dimensional with the length $T = 50000$ and was obtained from online\footnote{\url{http://research.ics.aalto.fi/ica/cocktail/cocktail\_en.cgi}} as in \cite{MiettinenNordhausenTaskinen2017}. To this we added 17 channels of standard normal noise to obtain the latent $\bm{z} = (s_1, s_2, s_3, w_4, w_5, \ldots, w_{20})^\top$ which was then mixed with a matrix $\bm \Omega \in \mathbb{R}^{20 \times 20}$ containing iid standard normal variables to obtain the ``observed'' data $\bm x _t = \bm \Omega \bm z _t$.

We considered all four bootstrap strategies using AMUSE with lag 1 and SOBI with lags $1 , \ldots , 12$. Each combination then produced a string of $p$-values $p_0, \ldots , p_{19}$ corresponding respectively to the null hypotheses $H_{0,0}, \ldots , H_{0,19}$. The forwards estimate for $d$ is then the first $k$ for which $H_{0, k}$ is not rejected and the backwards estimate for $d$ is $k + 1$ where $k$ is the last $H_{0, k}$ to be rejected. The resulting estimates are shown in Table \ref{table::sound} and reveal that all combinations correctly identify the true signal dimension. Note that as the forwards and backwards estimates yield the true dimension then also any divide-and-conquer methods are bound to find the true dimension in this case. 

\begin{table}
\centering
\caption{The estimates for $d$ for each combination of bootstrap strategy and methods in the sound example.}\label{table::sound}
\begin{tabular}{p{2cm}p{2cm}p{0.1cm}p{1.6cm}p{1.6cm}p{1.6cm}p{1.6cm}}
Estimator & BSS & & parametric & non-par I & non-par II & non-par III \\
\hline
Forwards & AMUSE & & 3 & 3 & 3 & 3  \\
Forwards & SOBI & & 3 & 3 & 3 & 3 \\
\hline
Backwards & AMUSE & & 3 & 3 & 3 & 3  \\
Backwards & SOBI & & 3 & 3 & 3 & 3  \\
\hline
\end{tabular}
\end{table}

\section{Summary}

We proposed four bootstrap tests to test the signal subspace dimension in an SOS framework using AMUSE or SOBI. Simulations showed that the different bootstrap tests work generally well and keep the $\alpha$-level with good rejection power. To estimate the subspace dimension, the tests would need to be applied sequentially, maybe with different strategies and a possible need for multiple testing adjustments. These issues will be addressed in future work, although an application to sound wave data already yielded some evidence that the sequential estimation works in practice. Note that the suggested tests ignore any possible variation coming from the estimation of the signal as these parts are not bootstrapped. Time series bootstrap strategies as described, for example, in \cite{Lahiri2003} could then be applied also here for the signal parts. This extension will also be explored in future research.
%
%

\section*{Acknowledgements}

The work of KN was supported by the CRoNoS COST Action IC1408.

\bibliographystyle{splncs}
\bibliography{refs}

\end{document}